\documentclass[a4paper,11pt]{article}
\usepackage{pos}
\usepackage{xspace}
\usepackage{xcolor}
\usepackage{framed}

\newcommand{\dphi}{d_\varphi}

\newcommand{\ORD}{{\cal O}}

\newcommand{\Lag}{{\cal L}}
\newcommand{\LagLO}{\Lag_{\text{LO}} }

\newcommand{\TEMT}{ \Tud{\rho}{\rho} }
\newcommand{\Tud}[2]{T^{#1}_{\;\; #2}}

\newcommand{\mink}{\eta}

\newcommand{\Op}{{\cal O}}

\newcommand{\MeV}{\,\mbox{MeV}}

\newcommand{\gast}{\ga_*}
\newcommand{\best}{\be_*}

\newcommand{\bestp}{\be'_*}

\newcommand{\matel}[3]{\langle #1|#2|#3\rangle}

\newcommand{\al}{\alpha}
\newcommand{\be}{\beta}
\newcommand{\ga}{\gamma}

\newcommand{\de}{\delta}
\newcommand{\De}{\Delta}

\newcommand{\sig}{ \sigma}

\newcommand{\FIG}{Fig.~}

\newcommand{\vev}[1]{\langle #1 \rangle} 
\newcommand{\state}[1]{|#1\rangle}
\newcommand{\astate}[1]{\langle #1|}

 \title{Soft Theorems and Dilaton Effective Theory }

\author*{Roman Zwicky}

\affiliation{Higgs Centre for Theoretical Physics, School of Physics and Astronomy, University of Edinburgh,\\
	Edinburgh EH9 3JZ, Scotland}

\emailAdd{roman.zwicky@ed.ac.uk}

\abstract{We derive a new model-independent double-soft dilaton theorem, taking into account the spacetime dependence of the dilation commutator 
$[i Q_D,{\cal O}(x)]= (\Delta_{\cal O} + x \cdot \partial){\cal O}(x)$. 
The procedure restores positivity in the  (pseudo)-Goldstone masses and 
sets the constraint  $\Delta_{\cal O} = d-2\,$  for a single operator ${\cal O}$ responsible for 
generating a dilaton mass.We discuss gravitational form factors as a tool to probe infrared conformality in field theories with particle content. In a second part we explore to what extent  QCD-like gauge theories (in the chiral limit) could fit into this category.
We find  that  the quark bilinear has scaling dimension $\Delta_{\bar qq} = d-2$,
 therefore satisfying the double-soft theorem. We show that some findings are realised in  
  ${\cal N}=1$   supersymmetric gauge theories and argue that the extension below the conformal window makes sense in that case.}

\FullConference{The 41st International Symposium on Lattice Field Theory (LATTICE2024)\\
 28 July - 3 August 2024\\
Liverpool, UK\\}

\tableofcontents

\begin{document}
\maketitle

\section{Introduction }

The dilaton in our context is  the Goldstone boson due to spontaneous scale 
symmetry breaking.  Originally introduced and developed to describe the 
strong interactions e.g. \cite{Isham:1970gz,Zumino,Ellis:1971sa},  
it has found numerous applications in particle physics and cosmology \cite{Cata:2018wzl,Rho:2021zwm,Zwicky:2023bzk,Zwicky:2023krx,Appelquist:2024koa}. 
Recent interest has been spurred by a light singlet $J^{PC} = 0^{++}$ state emerging 
in lattice simulations  at finite quark mass.  
We begin by  reviewing model-independent soft dilaton theorems in section \ref{sec:soft}, 
verify them in dilaton effective theory in section \ref{sec:DET}
before discussing the possibility that gauge theories 
in the chirally broken phase, including QCD,  are described by an 
infrared fixed  point (IRFP) with a dilaton. 
For the latter,  prominent features are the extraction of scaling dimensions and the impact on gravitational form factors. 

\section{Model-independent Soft Dilaton Theorem}
\label{sec:soft}

Soft theorems were part of the current algebra  arsenal  in the 1960's and assume that  
a (pseudo) Goldstone (pG) is much lighter than the remaining particles (or hadrons)  
$m_{pG} \ll m_{\text{had}}$. 
In practice, this amounts to replacing the soft Goldstone by a symmetry 
transformation.\footnote{In the language of correlation function this corresponds 
to the reduction of a $n$-point to a ($n$-$1$)-point function. For example, 
a scalar-vertex with zero momentum insertion can be replaced by mass-differentiation.} 
Formulated in the concrete case of  spontaneously broken scale invariance it  reads
 \begin{equation}
\label{eq:soft2}
 \lim_{q \to 0} \matel{D(q) \be}{{\cal O}(0)}{\al}  = - \frac{1}{F_D} \matel{ \be}{i [Q_D,{\cal O}(0)]}{\al}  \;+\;  \lim_{q \to 0} i q \cdot R 
\;+\;  \ORD ( m^2_D/m^2_{\text{had}} ) \;, 
\end{equation}
where  
   \begin{equation}
\label{eq:remainder}
  R_\mu =   - \frac{i}{F_D}\int d^d z e^{i q\cdot z} \matel{ \be}{T J^D_{\mu}(z) {\cal O}(0)}{\al}  
\;, 
\end{equation}
 is the   remainder which vanish unless  $\state{\al,\be}$ are degenerate with intermediate  states 
and $Q_D = \int d^{d-1}z J_0^D(z)$  the dilatation charge. The quantity 
$F_D$ is the dilaton decay constant and the \emph{order parameter} of the spontaneous scale symmetry breaking.  It is defined through the $T_{\mu\nu}$,
the energy-momentum tensor\footnote{It is related to the Goldstone current by
$J_D^\mu = x_\nu T^{\mu\nu}$ such that $\matel{0}{J^D_{\mu}}{D(q)} = i q_\mu F_D$.  
The version for pions  of \eqref{eq:soft2} for chiral symmetry breaking 
is obtained by replacing  
$F_D \to F_\pi$, $\astate{D} \to \astate{\pi^a}$ and $J^D_\mu \to J^{5a}_\mu = \bar q T^a \ga_\mu \ga_5 q$.}
\begin{equation}
\label{eq:FD}
\matel{0}{T_{\mu\nu}}{D(q)} =  
\frac{F_D }{ d-1}  
(m_D^2 \eta_{\mu\nu} - q_\mu q_\nu) \;,
\end{equation}
with $\eta_{\mu\nu} $
 (the mostly minus)  Minkowski metric. 
The novel element lies in  the application of the dilatation commutator  
to a primary operator 
\begin{equation}
\label{eq:comm}
i [Q_D ,\Op(x)] =  \frac{1} {F_D} (  \De_{\Op}  + x \cdot \partial) \Op (x)  \;,
\end{equation}
for which   $x \cdot \partial \equiv x_\mu \partial^\mu$ and $\De_\Op$ 
express the transformation of the argument and the operator respectively. 
  In \eqref{eq:comm}  $\De_\Op = d_\Op + \ga_\Op$ 
is known as the scaling dimension which differs from the engineering dimension by the anomalous part.  In order to appreciate this aspect we first proceed to repeat the textbook derivation of 
\eqref{eq:soft2} which starts by applying the Lehmann-Symmanzik-Zimmermann (LSZ)  procedure 
\begin{equation}
\label{eq:start}
\matel{D(q) \al}{\Op(0)}{\be}  =  \frac{ i}{Z_D} \lim_{q^2 \to m_D^2} ( m_D^2-q^2) \int 
d^d z  e^{iq \cdot z} 
\matel{ \al} {T \partial  \cdot J^D(z) \Op(0)}{\be}  \;,
\end{equation}
where $\matel{D}{ \partial  \cdot J^D}{0} = Z_D$ serves as the interpolating operator for the dilaton. Next one considers the standard Ward identity 
\begin{equation}
\label{eq:WI}
\partial^\mu \matel{\al} {T J^D_\mu(z) \Op(0)}{\be} = 
 \de(z_0) \matel{\al}{[J^D_0(z), \Op(0)]}{\be} + 
 \matel{\al} {T \partial \cdot J^D(z) \Op(0)}{\be}  \;,
\end{equation}
where the first term on the right corresponds to the symmetry transformation of the 
operator $\Op$ and the second one 
parameterises  possible explicit or anomalous symmetry breaking. In the case of the pion 
the latter corresponds to the partially conserved axial 
current proportional to the pion mass (explicit breaking).  We then substitute \eqref{eq:WI} 
into \eqref{eq:start}.
 Up to this point everything is exact. The soft theorem \emph{assumption} is that the matrix element in \eqref{eq:start} 
changes smoothly as the Goldstone is made soft $q \to 0$. Proceeding one obtains  
the soft theorem \eqref{eq:soft2} 
\begin{equation}
\label{eq:meaning}
\matel{D(q) \al}{\Op(0)}{\be}   =  \lim_{q \to 0} \matel{D(q) \be}{{\cal O}(0)}{\al}|_{\mbox{\eqref{eq:soft2}}}
+  \ORD ( m^2_D/m^2_{\text{had}} ) \;,
\end{equation}
with the commutator part originating from the first term 
in \eqref{eq:WI} and the remainder from the left-hand side of the same equation. The 
order $\ORD ( m^2_D/m^2_{\text{had}})$ is the minimal consequence of taking the soft limit. 
Note that Eq.~\eqref{eq:meaning} gives meaning to the statement of an ``off-shell matrix 
element". 
  We now proceed to apply 
the soft theorem using \eqref{eq:comm} to the two special cases of one and two soft dilatons: 
$\matel{D}{\Op}{0}$ and $\matel{D}{\Op}{D}$.
\paragraph{Single-soft theorem:} evaluating $\matel{D}{\Op}{0}$ is straightforward 
as the $x$-dependence in \eqref{eq:comm} drops out, one gets, with $\vev{\Op} \equiv \matel{0}{\Op}{0}$, 
\begin{equation}
\label{eq:single}
\matel{D}{\Op}{0} =  -\frac{1}{F_D} \De_{\Op} \vev{\Op} \;.
\end{equation}
\paragraph{Double-soft theorem:} the evaluation of 
 $\matel{D(q)}{\Op}{D(q')}$  is more subtle and one has to proceed step by step. Similarly,  we get
\begin{equation}
\matel{D(q)}{\Op(x)}{D(0)} =  -\frac{1}{F_D}( \De_{\Op}  + x \cdot \partial)  \matel{D(q)}{\Op(x)}{0} \;,
\end{equation}
but this time the derivative cannot be neglected since 
$ \matel{D(q)}{\Op(x)}{0} \propto e^{i q\cdot x}$ is $x$-dependent.  
To understand what to do one has to remember that operators have to be smeared 
$\Op_h = \int d^d x  \Op(x) h(x)$, 
by  test function  $h \in {\cal S}(\mathbb{R}^d)$ in Schwartz space  
(sufficiently fast fall off at infinity, e.g. $h(x) =  r^n e^{- |c| r^2}$). Taking this into account one finds 
$\matel{D}{\Op_h}{D} = 1/F_D^2 (\De_{\Op}-d)\De_\Op \vev{\Op_h}$, integrating by parts, 
and removing the test function this leads to our main formal result for the double-soft dilaton matrix element
\begin{equation}
\label{eq:double}
\boxed{\matel{D}{\Op}{D} =  \frac{1}{F_D^2}(\De_{\Op}-d)\De_\Op \vev{\Op} } \;.
\end{equation}
Whereas the single-soft theorem is  standard, the double-soft theorem \eqref{eq:double} is 
a new result  \cite{Zwicky:2023krx} to the best of our knowledge.  The derivation given here is more detailed than in that reference.

In order to further refine results we consider the following two fundamental formulae 
\begin{equation}
\label{eq:one}
F_D m_D^2 = \matel{D(q)}{\TEMT}{0}  \;, \quad 2 m_D^2 = \matel{D(q)}{\TEMT}{D(q)}  \;, 
\end{equation}
where the first one follows from \eqref{eq:FD} and the second one 
 follows from $P_\mu = \int d^{d-1}x T_{\mu0}$.   We may obtain valuable results 
 by assuming that a single irrelevant ($\De_{\Op} \leq d$) operator 
 $\Op \subset \TEMT$   is responsible for generating the mass via the matrix elements in \eqref{eq:one}.  First we note that  the vacuum expectation value must be  lower  than the perturbative vacuum  and thus negative  $\vev{\Op} < 0$.
   Applying the single- and double-soft theorem to \eqref{eq:one} one 
  immediately  gets
 \begin{equation}
\label{eq:De2}
 m_D^2 = - \frac{1}{F_D^2}  \De_\Op \vev{\Op} \;, \quad 
  m_D^2 =\frac{1}{2 F_D^2}  (\De_{\Op}-d)\De_\Op \vev{\Op} \;,
\end{equation}
from which the following conclusions can be drawn
 
 \begin{framed}
\begin{itemize} 
\item [A)] The $d$-term in \eqref{eq:De2} assures  $m_D^2 > 0$ since $\Op$ is irrelevant ($\De_\Op \leq d$). 
 \item [B)] The two equation in \eqref{eq:De2} hold simultaneously if and only if $\De_\Op = d-2 $.
\item [C)] Assuming $\De_\Op \neq d-2$, the only solution is 
$\vev{\Op} =0$.
\end{itemize}
\end{framed}
In summary the $d$-term resolves a positivity problem and the operator scaling dimension 
must be $d-2$ with the latter being a  result of practical importance. 

 At last two cautionary remarks. 
If $\TEMT \supset \sum_i \Op_i$ with $\De_{\Op_i} \neq \De_{\Op_j}$ are contributing to  \eqref{eq:one}, then no such strong statement can be made. 
This is for example the case in the Gross-Neveu-Yukawa theory \cite{Cresswell-Hogg:2025kvr} in $d=3$ where the $\varphi^3$ and the Yukawa-operator take on these roles. 
In the case where there are other Goldstone such as the pion with $m_D > 2 m_\pi$, the dilaton becomes unstable and one must involve a more elaborate formalism. 
Nevertheless, 
we may expect that the qualitative results still hold. 

\section{Dilaton Effective Theory}
\label{sec:DET}

Here, we consider the dilaton  on its own without further Goldstone bosons such as the pions. 
The latter are readily added and are compatible with the pions, or even better,   resolve the Goldstone improvement problem \cite{Zwicky:2023fay}.  
The leading order dilaton Lagrangian  reads
\begin{equation}
\label{eq:Leff}
\LagLO =  \frac{1}{2}  (\partial  \chi)^2 -  \frac{\xi_d}{2} \,  
 R \,  {\chi}^{2}  + 
V (\hat{\chi})   \;, 
\quad \xi_d \equiv \frac{\dphi}{2(d-1)} \;, 
\end{equation}
with $\dphi \equiv \frac{d-2}{2}$ the dimension of the free scalar and coset fields ($\mathbb{R}^{d} \to \mathbb{R}_+$)
\begin{equation}
 \hat{\chi} \equiv  e^{-\frac{D}{F_D}} \;, \quad 
\chi  \equiv  (  F_D/\dphi )  \hat{\chi}^{\dphi}   \;, \quad
\end{equation}
convenient for the effective theory.
The first two terms in \eqref{eq:Leff} are separately invariant under global Weyl transformations
\begin{equation}
g_{\mu\nu} \to e^{-2 \al} g_{\mu\nu} \;, \quad  D \to D -  \al F_D \;, \quad \Rightarrow 
\quad \hat \chi  \to e^{\al} \hat  \chi \;, \quad \chi  \to e^{\al \dphi} \chi \;, 
\end{equation}
they are invariant together under \emph{local}  Weyl transformations. 
This is the essence for the celebrated improvement term \cite{Callan:1970ze}, 
see \cite{Zwicky:2023fay} for the dilaton context and further references.  
  The exponential representation
   is said to realise the symmetry non-linearly, which we will verify explicitly later on.
Choosing the $F_D > 0$ phase convention, then implies $\chi > 0$.
Since the improvement term  assures local conformal symmetry it gives rise to 
characteristic dilaton physics:  
it  realises the fundamental Goldstone matrix element \eqref{eq:FD} 
in the effective theory \emph{and} affects gravitational form factors decisively  (as discussed later on). 

\subsection{A quick note on a dilaton potential}

Turning to the potential   
$V (\hat{\chi})$, we may make a surprising connection with the double-soft theorem.  Firstly, 
for a massless dilaton the  potential vanishes $V (\hat{\chi}) = 0$, 
and when a dilaton mass  is generated by 
a single operator $\Op$, as in the previous section, then it assumes the form 
\begin{equation}
\label{eq:V}
 V_{\De_\Op} (\hat{\chi}) =   \frac{m_D^2 F_D^2 }{ \De_\Op -d} \left( \frac{1}{\De_\Op} \hat{\chi}^{\De_\Op} - \frac{1}{d}  \hat{ \chi}^d \right) =  c_V  +   \frac{1}{2} m_D^2    {D}^2  + \ORD(D^3) \;,
\end{equation}
with $c_V$ a constant of no concern to us.
A long time ago,  it was observed by Zumino \cite{Zumino}
 that $V_{\text{Zumino}} \propto \hat{ \chi}^d $ 
 is the only term allowed by scale invariance itself.  However, in that case the potential has no minimum and thus one 
 needs to add at least one more operator  $V = V_{\text{Zumino}} + c \hat{\chi}^\De$, which is one way to motivate  \eqref{eq:V}.   
The new observation here is that the role of the Zumino-term, in the context of the double-soft theorem, is taken on by $x \cdot \partial$-term as it leads  to the $d$-factor  in 
the second equation in  \eqref{eq:De2}, since 
$\matel{D}{\hat{\chi}^\De - \frac{\De}{d}  \hat{ \chi}^d }{D} = \frac{1}{2} \De (\De-d)$. 
This further enhances confidence in the consistency of our framework.
  
\subsection{Explicit commutators in the effective theory}

We find it instructive to realise the fundamental commutator \eqref{eq:comm} in the 
effective theory for the   operator 
\begin{equation}
\label{eq:not}
\Op = \hat{\chi}^{\De_{\Op}} = f(\hat{D}) \;,  \quad \hat{D} \equiv \frac{D}{F_D} \;,
\end{equation}
of scaling dimension $\De_{\Op}$, that is $\Op \to e^{\al \De_\Op }\Op$ under a Weyl transformation. Thus the question how to compute 
a commutator. One may either use the Bjorken-Johnson-Low formula or 
proceed by integrating  the Ward identity \eqref{eq:WI} to obtain 
\begin{equation}
\label{eq:com}
\vev{{i [Q_D ,\Op(x)]}} = - i  \int d^d z    \vev{T  \partial \cdot  J^D (z) \Op(x)} \;,
\end{equation}  
assuming the  vanishing of the correlator  at infinity. 
The divergence of the dilatation current $J_D^\mu(z) = z_\nu T^{\mu\nu}(z)$ is 
\begin{equation}
\label{eq:partial}
\partial \cdot J^D = \TEMT + z_\nu \partial_\mu T^{\mu\nu} \;,
\end{equation}
 and one  anticipates that the first and the second term are responsible for the scaling dimension and the $x \cdot \partial$-term in \eqref{eq:comm} respectively.   
 To progress concretely we need 
the energy-momentum tensor which we may obtain by metric variation 
 \begin{alignat}{2}
&T_{\mu \nu}  &\;=\;& -2  \frac{ \de  }{  \de{g^{\mu\nu}}} \int d^d x  \sqrt{-g} \LagLO\big|_{g_{\mu\nu} = \mink_{\mu\nu} }  \nonumber \\[0.1cm]
 & &\;=\;& 
        \partial_\mu \chi  \partial_\nu \chi   - g_{\mu\nu} (  \frac{1}{2} (\partial\chi)^2  - V(\hat{\chi}))    + 
       \xi_d   (g_{\mu\nu} \partial^2 -  \partial_{\mu} \partial_{\nu} )     \chi^{2}  \;.
   \end{alignat}
The individual terms in \eqref{eq:partial} evaluate to 
\begin{alignat}{4}
\label{eq:TT}
&   \TEMT  &\;=\;&  \dphi\,  {\chi}   \partial^2 {\chi}  + d \, V &\;=\;&   
-   F_D \partial^2{D} &\;+\;& \dots   \;,  \nonumber \\[0.1cm]
&  \partial_\mu T^{\mu\nu}  &\;=\;&  \partial^\nu {\chi}  \partial^2 {\chi} + \partial^\nu V(\hat{\chi})  &\;=\;&   
+  (\partial^\nu  {D}) \partial^2 {D} &\;+\;&  \dots  
    \;, 
\end{alignat}
where higher order terms in the dilaton field  and the potential are hidden in the dots in the second equality.  The potential terms would give rise to the $\ORD ( m^2_D/m^2_{\text{had}} )$ corrections which necessitate 
a more elaborate analysis. 
In order  to evaluate \eqref{eq:com} further we use 
\begin{equation}
 F_D \partial^2 \vev{T D(z) f(\hat{D}(x))} = -  i\de^{(d)}(z-x) \vev{ f'(\hat{D}(x)) } \;,
\end{equation}
 which follows from $\partial^2 \vev{T D(z) D(x)} = -  i\de^{(d)}(z-x)$,   We may now infer 
\begin{alignat}{3}
&   \int  d^d z   \vev{T \TEMT(z)  f(\hat{D}(x)) }   &\;=\;&     - i \vev{f'(\hat{D}(x))  }   &\;+\;&  \dots
  \;,
    \nonumber \\[0.1cm]
&   \int  d^d z   \vev{T   z_\nu \partial_\mu T^{\mu\nu}(z)  f(\hat{D}(x)) }   &\;=\;& 
 -  i \vev{x \cdot \partial \hat{D}(x)  f'(\hat{D}(x)) }   &\;+\;&   \dots   \;,
\end{alignat}
where the dots have the same meaning as above.  Assembling bits and pieces we get
 \begin{alignat}{3}
& \vev{i [Q_D ,f(\hat{D}(x))] }   &\;=\;& \vev{(1+x \cdot \partial \hat{D})(-f'(\hat{D}(x))) }     
&\;+\;& \dots \nonumber \\[0.1cm]
 & &\;=\;& 
\vev{(\De_{\Op} + x \cdot \partial) f(\hat{D}(x)) } &\;+\;& \dots
   \end{alignat}
where the explicit function form in \eqref{eq:not} has been made use of.  We have therefore explicitly 
computed the  commutator \eqref{eq:comm} in the effective theory. 
Note that, for non-integer $\De_{\Op}$ 
 the derivation  goes beyond the free field theory.\footnote{For a  non-primary operator (e.g. 
 a descendant) the fundamental commutator \eqref{eq:comm} is altered.  
 For a free field theory where $E = (\partial^2+m_{\varphi}^2)   \varphi^2$ is an equation of motion operator, $\vev{[Q_D,E(x)]} = 0$ must hold 
 which we have verified explicitly.}  

\paragraph{The case Pions:}
It is instructive to consider Goldstone bosons arising from the spontaneous breaking of an internal symmetry.
As a concrete example, we have in mind the pions, which emerge from the spontaneous breaking of the flavor symmetry
$SU(N_f)L \times SU(N_f)R \to SU(2)V$ down to the isospin subgroup.
To the effective Lagrangian one may add the following Weyl-invariant kinetic term
\begin{equation}
\label{eq:kin}
\de \Lag = \frac{F_\pi^2}{4} \hat{\chi}^{d-2} \text{Tr} [ \partial_\mu U \partial^\mu U^\dagger] 
\end{equation}
where $U = \exp( i \pi/F_\pi)$ is the coset field. The pion has zero Weyl-weight vanishes, as can be shown directly from the conformal algebra~\cite{Ellis:1971sa}, consistent with the interpretation of $\pi$ as a generalised angle.
Thus we must have
\begin{equation}
\label{eq:pi}
[i Q_D, \pi] =  x \cdot \partial \pi \;,
\end{equation}
which we would like to verify by computing the leading order  commutator.   
For simplicity, both here and above, we suppress the pion’s flavor index, since it plays no special role and can be restored trivially.
The relevant contributions from the kinetic term, analogous to those in \eqref{eq:TT}, are
\begin{equation}
 \TEMT = \dphi(  {\chi}   \partial^2 {\chi} - (\partial \pi)^2 )  \;, \quad 
\partial_\mu T^{\mu\nu} =   (\partial^\nu  {D}) \partial^2 {D}  + \dots \;.
\end{equation}
In the very same way as for the dilaton the second term will generate the translation in \eqref{eq:pi}. 
Let us therefore focus on the Weyl-weight part in \eqref{eq:com}
\begin{equation}
[i Q_D, \pi] |_{\TEMT}  =   - i  \int d^d z    \vev{T     \TEMT (z) \pi(x)} 
=  A + B  \;,
\end{equation}
where  $B$
\begin{equation}
B =   i  \dphi  \int d^d z    \vev{T  (\partial \pi)^2 )  (z) \pi(x)}  \;,
\end{equation}
is the second term, while the first term can be transformed into 
\begin{alignat}{2}
&  A &\;=\;&  - i  \dphi  \int d^d z    \vev{T     {\chi}   \partial^2 {\chi}   (z) \pi(x)}  =
- i  F_D   \int d^d z \vev{T  \partial^2 D(z)  \pi(x)} \nonumber \\[0.1cm]
&   &\;=\;&  
- i  F_D   \int d^d z \vev{T  \partial^2 D(z) (i \int d^d y \frac{1}{2} {\hat{\chi}^{d-2}} (\partial \pi)^2)(y)   \pi(x)}  =  - B \;,
\end{alignat}
upon using leading-order perturbation theory and expanding $\hat{\chi}^{d-2} \to \dphi D(y)/F_D$.
Thus, the two contributions cancel, yielding $[i Q_D, \pi]|_{\TEMT} = 0$, which confirms Eq.~\eqref{eq:pi} and establishes that the pion indeed has vanishing Weyl weight, $\De_\pi = 0$.
Of course, this aspect is instilled in the way that the dilaton acts as a compensator in \eqref{eq:kin} 
but it is nevertheless instructive to see how it works out.

\section{The Deep Infrared}
\label{sec:DIR}

Let us denote by $\sig$ and $\varphi$
the lightest  state with vacuum quantum numbers 
$J^{PC} = 0^{++}$   and a generic massive 
state, respectively. Note that 
all states, except the $\sig$ and other Goldstones, are expected to be massive.  
There are then three logical possibilities regarding the mass hierarchies 
(where we work in the limit of no explicit scale symmetry breaking):\footnote{In the gauge theory, this corresponds to setting the quark masses to zero (chiral limit).}
 \begin{framed}
 \vspace{-0.6cm}
 \begin{alignat*}{4}
& (a) &  & \;\;  \text{  The $\sig$ remains heavy -  {\bf no dilaton}: }  &  &  \bar{m}_{\sig}  = \ORD(1)\bar{m}_\varphi   \;\; &-&  \;\; \text{ soft theorems do not apply} 
\nonumber \\[0.1cm]
& (b) &  & \;\;  \text{  The dilaton as {\bf pseudo Goldstone}: }  \;\; &  &  \bar{m}_\sig \ll \bar{m}_\varphi &-&  \;\; \text{ soft theorems are approximate} 
\nonumber \\[0.1cm]
& (c) &  & \;\;  \text{  The dilaton as a {\bf genuine  Goldstone}: }&  &  \bar{m}_\sig =0  &-&  \;\; \text{ soft theorems become exact} 
\end{alignat*}
\vspace{-0.9cm}
\end{framed}
There is consensus in the community that  case (c) occurs when 
a conformal field  theory (CFT) is spontaneously broken.  In  contrast, in the presence of a 
renormalisation group (RG) flow, such as in QCD, (a) is  the dominant expectation,  
 although possibility (b) is sometimes considered. In fact, it is not easy to make sense out of 
 case (b), for a serious attempt in QCD see \cite{Golterman:2024vra}.
As for case (c), one might wonder  whether  a RG flow and a massless dilaton are compatible at all \cite{Zwicky:2023krx}.  Colloquially speaking: ``Does the Goldstone candidate remember that the spontaneously broken scale symmetry is only emergent?"   The answer to this question, or its necessary refinements, are 
not known in general.  
However,  at least the previously mentioned Gross-Neveu-Yukawa theory in $d=3$ provides an example realising scenario (c)  \cite{Cresswell-Hogg:2025kvr}.   
The theory is asymptotically free, and a  (one-sided) flat quantum potential triggers spontaneous scale breaking through a vacuum expectation value.\footnote{The expectation value can be understood as a double-scaling limit of couplings assuming 
their critical values \cite{Cresswell-Hogg:2025kvr}. 
In addition, the double-soft theorems \eqref{eq:double} work exactly as they should in this theory. } 
All dilaton parameters are calculable in terms of the latter. 
  Summary:
 \begin{framed}
 \vspace{-0.6cm}
 \begin{alignat*}{4}
& (1) &  & \;\;  \text{  Spontaneously broken CFT: }  &  &  \bar{m}_{\sig}  = 0  \;\; &-&  \;\; \text{ formally true} 
\nonumber \\[0.1cm]
& (2) &  & \;\;  \text{  RG flow spontaneous scale breaking: }  \;\; &  & 
 \bar{m}_{\sig}  = \ORD(1)\bar{m}_\varphi     \;\; &-&  \;\; \text{ dominant expectation in literature} 
\nonumber \\[0.1cm]
& (2')  &  &   &  &  \bar{m}_\sig \ll \bar{m}_\varphi  \;\;   &-&  \;\; \text{ sometimes considered}
\nonumber \\[0.1cm] 
& (2'')  &  &   &  &  \bar{m}_\sig =0  \;\;   &-&  \;\; \text{ GNY theory as an example \cite{Cresswell-Hogg:2025kvr} } 
\end{alignat*}
\vspace{-0.9cm}
\end{framed}

\subsection{Massless dilaton and gravitational form factors}
\label{sec:grav}

Gravitational form factors offer an interesting setting to test these ideas, 
since the momentum transfer resolves the theory at different scales.  These form factors
are  analogous to usual form factors, with  the 
energy-momentum tensor playing the role of the transition current.  
We refer to  \cite{Polyakov:2018zvc} for a review on this flourishing subject. 
In particular since they are partially measurable in experiment   \cite{Burkert:2018bqq,Duran:2022xag}  they have been investigated by many methods (see \cite{Polyakov:2018zvc} for references).

We will consider a form factor of a scalar particle $\varphi$ for simplicity.  
The standard definition is (except using a calligraphic D to avoid confusion with the dilaton itself)
\begin{equation}
\label{eq:GFF}
\Theta_{\mu\nu}(q)  \equiv  \matel{\varphi(p')}{T_{\mu\nu}(0)}{\varphi(p)} = 
   2 {\cal P}_\mu {\cal P}_\nu A (q^2) +
\frac{1}{2} (q_\mu q_\nu-  q^2 \mink _{\mu\nu} ) {\cal D} (q^2)  \;,
\end{equation}
where $q \equiv p'-p$ and $  {\cal P} \equiv 
\frac{1}{2}(p + p')$ are the 
 momentum transfer  and momentum average, respectively.  
 The Lorentz structures  obey $q^\mu \Theta_{\mu\nu} = 0$ ensuring   translational invariance. Furthermore,   one has $A(0) =1$ in all generality, since $T_{\mu\nu}$ is the Noether current of translation,  and thus
 $P_\mu = \int d^{d-1} x T_{\mu 0}$.  Now, if ${\cal D} (q^2)$ is finite for $q \to 0$ one has the well-known textbook formula $\Theta^{\rho}_{\phantom{\rho} \rho}(0) =2 m_\varphi^2$ \cite{Donoghue:2022wrw} 
   (see also  \eqref{eq:one}). 
How will IR conformality manifest itself? Formally, by the Ward identities, one expects 
\emph{IR-conformality}:
\begin{equation}
\Theta^{\rho}_{\phantom{\rho} \rho}(0) = \matel{\varphi(p)}{\TEMT}{\varphi(p)} =0 \;,
\end{equation} 
in the case of an IRFP.  This is at odds with the above-mentioned  textbook formula.  The caveat is that the smoothness assumption in the ${\cal D}$ form factor is not true in the case of a massless dilaton, which gives rise to a pole singularity  
\begin{equation}
\label{eq:D}
{\cal D} (q^2) =  \frac{4 }{d-1} \frac{m_\varphi^2}{q^2} + \ORD(1) \quad \Rightarrow  \quad 
\Theta^{\rho}_{\phantom{\rho} \rho}(0) = 0 \;,
\end{equation}
and IR-conformality,  
as can easily be inferred from \eqref{eq:GFF}.
An explicit computation,  
using the LSZ procedure  \cite{DelDebbio:2021xwu}  or the effective theory  \cite{Zwicky:2023fay}, yields
\begin{equation}
{\cal D} (q^2) = \frac{4}{d-1}\frac{ g_{D\varphi \varphi} F_D}{q^2} + \dots
\end{equation}
where  the couplings is defined by 
$ \Lag_{\text{eff}}  \supset \frac{1}{2} g_{D \phi\phi} D \varphi \varphi$. 
Using the compensator formalism, it is readily established that 
\begin{equation}
\label{eq:GT}
g_{D \phi\phi} = \frac{2 m_\varphi^2}{F_D}  \;.
\end{equation} 
Colloquially speaking, 
\emph{``the role of the dilaton is implement the conformal Ward identity.``}\footnote{This is analogous to QCD where the pion implements the chiral Ward identity via 
the  Goldberger-Treiman mechanism   $g_{\pi \varphi\varphi} = 2 m_\varphi^2/F_\pi$ 
(primarily known when $\varphi$ is the nucleon). 
The analogy to \eqref{eq:GT} needs no further explanation.}

In summary one has:
  \begin{framed}
 \vspace{-0.6cm}
 \begin{alignat*}{5}
&   \text{  No dilaton pole: }   \;\;  \;\;  \;\;     & & 
 {\cal D}(q^2) = c_0 &+\; &   \ORD( q^2)   &  &  \;\;  \;\;  \;\;  \matel{\varphi(p)}{\TEMT}{\varphi(p)}  = 2m_\varphi^2 & &   \;\;  \;\;  \;\;    \text{ textbook formula} 
\nonumber \\[0.1cm]
&  \text{  A dilaton pole: }   \;\;  \;\;  \;\;    & & 
 {\cal D}(q^2) = \frac{c_{-1}}{q^2} &+\; &  \ORD(1) &  &    \;\;  \;\;  \;\;    \matel{\varphi(p)}{\TEMT}{\varphi(p)}  = 0 & &    \;\;  \;\;  \;\;  \text{ infrared conformality} 
\end{alignat*}
\vspace{-0.7cm}
\end{framed}
The cases with  explicit symmetry breaking  for the scalar, fermion, pion and dilaton 
gravitational form factors are considered in  \cite{Zwicky:2023fay},  and an analysis of the 
lattice results of the QCD gravitational form factors is to follow \cite{prep}.

\section{Chirally broken Gauge Theories with an IRFP}

Following the model-independent part, we now consider the possibility that gauge 
theories can be described by an IRFP.  For that purpose, 
it is instructive to first give an executive summary 
of the conformal window of gauge theories. 

\subsection{Conformal window and some facts about the $\sig$-meson}

We may consider 
 $N_f$ massless quarks in some representation of a gauge group $G$. 
The general study of the different phases in the $N_f$-$N_c$-plane is known as the conformal window, 
whose study was prompted by (walking) technicolor \cite{Hill:2002ap,Cacciapaglia:2020kgq}
and Seiberg dualities 
in ${\cal N}=1$ supersymmetric gauge theories \cite{Intriligator:1995au,Terning:2006bq,Shifman:2012zz}.
For concreteness, we  consider the non-supersymmetric case with 
$G = SU(N_c=3)$ and  quarks in the  fundamental representation. 
What is well established is the following: for $N_F > 16$, asymptotic freedom is lost, and below the theory develops a weakly coupled Caswell-Banks-Zaks fixed point. As 
 $N_f$ is lowered, the theory becomes more strongly coupled in the IR,  and at some critical 
  $N_f  = \bar{N}_f$  chiral symmetry breaking occurs.\footnote{The value of $\bar{N}_f$ is unknown, but there are indications
that it could be close to eight flavours.}  
The community's view is that for $N_f$ just below the critical value, scenario (b) might apply, whereas 
for $N_f = 3$ (which we refer to as QCD, since the light quark masses  $m_u + m_d \approx 10 \MeV$ and $m_s \approx 100 \MeV$ 
are relatively light), scenario (a) is the case.  

In nature, that is for $N_f = 3$ with the quark masses as above,  
the dilaton candidate is the  $f_0(500)$, known as the $\sig$-meson, which  
  continues to inspire particle physicists \cite{Pelaez:2015qba}, 
it decays quickly into two pions, defies the Regge trajectory  and large $N_c$ counting. 
Its rapid disintegration into two pions manifests itself in pole position siting deep in the complex plane  
$\sqrt{{s_{\sig}}}= m_{\sig} - \frac{i}{2} \Gamma_{\sig} =
 (400-550)  - i (200- 350)  \MeV $ \cite{ParticleDataGroup:2024cfk}.\footnote{The uncertainty is  conservative compared to $\sqrt{s_\sig} \!=\!  (441^{+16}_{-8} \!-\!i272 ^{+9}_{-12.5})   \MeV$
 \cite{Caprini:2005zr},  seen as having settled its existence.} 
Unlike the pion, the $\sig$-meson is not only sensitive to the up and down quark masses but also the strange 
quark mass (assuming the charm to be decoupled), just like the pseudoscalars
$\pi,K$ and $\eta$ gives a reasonable  $SU(3)_F$-octet description. 
What happens to the $\sig$-pole in the  $m_{u,d,s} \to 0$ limit is  generally unknown. 
As previously stated, the general expectation is that it won't change too much  \cite{Pelaez:2015qba}. 
However, there are some studies suggesting otherwise: 
\begin{itemize}
\item The $N_f = 4$ \cite{LatKMI:2025kti} lattice simulation with degenerate quarks 
 supports that  the $\sig$, which is stable here,  becomes rather light or even massless (see \FIG 13,14 in that paper). 
The ratio of $x \equiv (m_\pi/F_\pi)/(m_\pi/F_\pi)_{QCD} \approx 2,2.5,3$, serving as a measure for chirality,  for the respective data points is reasonably low.\footnote{Interest in 
dilaton interpretations have been spurred  by earlier lattice simulation, e.g. \cite{Ingoldby:2023mtf}, 
findings light ``$\sig$-mesons".}
\item The study by Oller \cite{Oller:2003vf}, using the phenomenological N/D-method and leading order chiral perturbation theory 
input, finds that the $\sig$ becomes stable at the $SU(3)_F$ symmetry point at $x \approx 2.5$, qualitatively similar to 
the lattice computation. 
 \item Fits of lattice data to gravitational form factors  \cite{prep}, indicate that QCD is not 
 incompatible with the dilaton Goldberger-Treiman mechanism. 
\end{itemize}
Arguments against a light $\sig$ mainly stem from its similar role to the $\kappa$-meson 
in phenomenology and  the inverse amplitude method  \cite{Pelaez:2015qba}. 
While these discussions  are tentative, more studies  with lighter quarks would be helpful 
in examining  the  $\sig$-pole in scattering amplitudes as well as  
 gravitational form factors.

A massless $\sig$ in the chiral limit, that is  case (c), is attractive as then 
  powerful statements can be made. This include exact results from soft theorems and gra 
 gravitational form factors.   
However, case (b)  might be equally interesting  as then  the Higgs can be the dilaton of a new gauge sector 
 provided that $F_\sig/F_\pi \approx 1$ (\cite{Zwicky:2023krx} and references therein).

\subsection{The pion sector - disregarding the dilaton}

Assuming scenario (b), where the $\sig$ remains massive in the chiral limit, we may consider the  pions  on their own.
In the deep IR, the massless pions 
correspond to a free field theory, which is particular fixed point.\footnote{This is for instance assumed when applying the a-theorem to QCD. Albeit there are subtleties with regards to conformality without the massless dilaton e.g.  \cite{Zwicky:2023fay}.}  
Can anything be learned in that case? Yes, we can extract scaling dimensions 
by matching the IRFP-assumption with chiral perturbation  theory ($\chi$PT) in the deep IR  \cite{Zwicky:2023bzk}. \emph{``The effective theory in its range of validity must describe the full theory."} For example,  for the scalar non-singlet operator  
$S^a = \bar q T^a q$
\begin{equation}
\label{eq:matchSa}
 \vev{S^a(x) S^a(0) }_{\text{IRFP}} =   \vev{S^a(x) S^a(0) }_{\chi\text{PT}} \;, \quad \text{for } x^2 \to \infty \;,
\end{equation}
with   scaling dimension  is $\De_{S^a} = (d-1) - \gast  $, where $\gast = \ga_m|_{\mu = 0}$ is  the quark anomalous dimension.  To evaluate the left-hand side 
we use that for a CFT one has
$\vev{\Op(x) \Op^\dagger(0) }_{\text{CFT}} 
\propto (x^2)^{-\De_\Op}$.\footnote{For more careful arguments with regards to the equation above see \cite{Zwicky:2023krx}, 
 the pseudoscalar non-singlet case for example. }  
The right-hand side can be dealt with source theory:
$S^a |_{\text{LO}} \propto  d^{abc} \pi^b \pi^c +  \ORD(1/F_\pi^2) $
such that the $\chi$PT-evaluation  reads $ \vev{S^a(x) S^a(0) }_{\chi\text{PT}} \propto 1/x^4$. Matching as in \eqref{eq:matchSa},  gives 
\begin{equation}
\label{eq:gastone}
 \vev{S^a(x) S^a(0) }_{\text{IRFP}} \propto \frac{1}{(x^2)^{3-\gast}}  \propto \frac{1}{x^4}
 \propto  \vev{S^a(x) S^a(0) }_{\chi\text{PT}}
   \quad \Leftrightarrow \quad \boxed{ \gast = 1} \;,
\end{equation}
a quark mass anomalous dimension of unity at the IRFP. 
This  is supported by alternative derivations, including hyperscaling arguments, 
the Feynman-Hellmann theorem, and the trace anomaly applied to the pions
\cite{Zwicky:2023bzk},
the decoupling of the dilaton for gravitational form factors \cite{Zwicky:2023fay} and 
the other correlators \cite{Zwicky:2023krx}.  Amongst those
 the Feynman-Hellmann-type argument is, perhaps, the strongest. 
Most notably, the result \eqref{eq:gastone} is consistent with 
 lattice simulations, perturbation theory and many model computations (see  \cite{Zwicky:2023bzk} for further references).
  
 Again, the result connects nicely to the soft theorem in that we may consider 
 the perturbation $\TEMT = (1+ \gast) N_f m_q \bar qq$ in \eqref{eq:De2}
 and observe that $\De_{\bar qq} = d-2$ does indeed hold, since $\De_{\bar qq} = \De_{S^a}$,
  and is therefore a consistent 
 perturbation  \cite{Zwicky:2023krx}; or conversely can be seen as another determination 
 of $\gast = 1$.    Such relations are known as dilaton Gell-Mann-Oakes-Renner relations in this context.  
      
 Another scaling dimension is $\De_{G^2} = d + \beta'_*$, 
 the one of the field strength tensor $G^2$, for which  $\bestp=0$ 
 is found from renormalisation group arguments and matching correlators  \cite{Zwicky:2023krx}.  Besides other advantages this makes it plausible that the trace anomaly emerges in the effective theory though loop computations rather than a Wess-Zumino-Witten type-term, which has to be put by hand to match the chiral anomaly in $\chi$PT.
 Moreover, $\De_{G^2}  =d$ 
 means that one must be cautious in applying partially conserved dilatation current formulae, since equation  \eqref{eq:one} are contradictory unless $\vev{\be/g G^2}_{m_q \to 0} =0$ 
 \cite{Zwicky:2023krx}.

 \subsection{Consistency with ${\cal N} =1$ supersymmetric gauge theories}
 
A key feature of ${\cal N}=1$ gauge theories is Seiberg duality, which states that
 an electric $SU(N_c)$ gauge theory is infrared dual to a magnetic $SU(N_f - N_c)$ gauge theory with an additional neutral ``meson field", as for instance reviewed in \cite{Intriligator:1995au,Terning:2006bq,Shifman:2012zz}.  
 A conformal window extends from $3 N_c \geq  N_f \geq \frac{3 }{2}N_c$. 
  From the NSVZ $\be$-function, it follows 
 that $\ga_*^{\text{el}} + \ga_*^{\text{mag}} = 1$,   impling that 
 at the lower end of the conformal window ($N_f = \frac{3}{2}N_c$)  one has
  $\gast^{\text{el}}  =1$ since $\ga_*^{\text{mag}}= 0$ is the weak coupling limit.  We notice in particular that this is consistent with 
  \eqref{eq:gastone} as previously claimed.  
   Using $\De_{\TEMT} = \De_{G^2} = 4 + \bestp$ and 
 matching correlation functions of the trace of the 
 energy-momentum tensor 
 \begin{equation}
\frac{1}{(x^2)^{4+ \best^{'\text{el}} }} \propto   \vev{\TEMT(x) \TEMT(0)}_{\text{el}} \propto 
 \vev{\TEMT(x) \TEMT(0)}_{\text{mag}}  \propto   \frac{1}{(x^2)^{4+ \best^{'\text{mag}}}} 
  \quad \Leftrightarrow \quad  
  \boxed{ \best^{'\text{el}}  = \best^{'\text{mag}} }\;,  
 \end{equation}
 one infers that the slope of the beta function in the electric and magnetic theories must be equal. 
 As above, one has $\best^{'\text{mag}}  =0$ at  the lower end of the conformal window 
 and thus  $\best^{'\text{el}}  =0$  \cite{Shifman:2023jqn}.  
 Again, this is consistent with the previous argument without supersymmetry. 
 
 More precisely we have shown that $\gast = 1$ holds at the end of  the conformal window. 
 One might wonder whether 
 the extension of $\gast = 1$ into the chirally broken phase itself makes any sense.  Seiberg duality seems to suggest that this is the case since the squark-antisquark bilinear $\tilde Q  Q$  in the electric theory is matched to the free meson $M$ in the magnetic dual \cite{Zwicky:2023krx}. Concretely,  we have  
 ($\De_{\tilde Q  Q} = 2 - \gast$)
 \begin{equation}
\frac{1}{(x^2)^{\De_M}} \propto  \vev{M(x) M(0)} \propto  \vev{\tilde Q  Q(x) \tilde Q  Q(0)} 
 \propto \frac{1}{(x^2)^{2 - \gast}}  \quad \Leftrightarrow \quad 
\gast =  2 - \De_M \;,
 \end{equation}
 implying that  $\gast =1$ must hold 
 in the IR-free regime $ N_c+1  \leq  N_f  \leq  \frac{3}{2} N_c$ since the \emph{free} meson's scaling dimension must be  $\De_M = 1$ . 
The Seiberg dualities reinforce the notion that the anomalous dimension $\gast = 1$ 
is robust below the conformal window, at least for  
  ${\cal N}=1$ supersymmetry.

\section{Conclusions}

In the first part, we focused on a model-independent discussion in the context of a dilaton (with refinements at the beginning of section \ref{sec:DIR}). In the second part, we explored whether gauge theories in the chirally broken phase can be interpreted in terms of an IRFP. 
The consistency of the results might indicate that this interpretation holds true in some gauge theories, 
such as, in decreasing likelihood, ${\cal N}=1$ supersymmetric theories,  and, non-supersymmetric gauge theories near or below the lower edge of the conformal window.
Here, we have focused only on a few essential aspects and emphasise that these proceedings should not be considered a review with a nearly complete set of references. It is clear that lattice QCD can play an important role in determining whether or not gauge theories have more to do with IRFPs than is generally assumed.

\paragraph{Acknowledgments:} 
RZ is supported by the
 STFC Consolidated Grant, ST/P0000630/1 and is grateful to Roy Stegeman for comments 
 on the draft.

\end{document}